\documentclass[12pt]{iopart}
\usepackage{cite}
\usepackage{graphicx}
\DeclareGraphicsExtensions{.eps}
\usepackage{bm}
\usepackage{iopams}
\usepackage{mathptmx}
\usepackage{textcomp}	 

\renewcommand{\d}{\mathrm{d}}
\renewcommand{\i}{\mathrm{i}}
\renewcommand{\Re}{\mathop{\mathrm{Re}}}

\renewcommand{\L}{\mathrm{L}}
\newcommand{\R}{\mathrm{R}}
\newcommand{\bra}[1]{\langle #1|}
\newcommand{\ket}[1]{|#1 \rangle}
\newcommand{\exponent}[1]{\ensuremath{\mathrm e^{#1}}}

\begin{document}

\title{Molecular electronics in junctions with energy disorder}

\author{Franz J Kaiser, Peter H{\"a}nggi and Sigmund Kohler}
\address{Theoretische Physik I, Institut
f{\"u}r Physik der Universit{\"a}t Augsburg, Universit{\"a}tsstr. 1, 86135
Augsburg, Germany} 
\ead{franz.josef.kaiser@physik.uni-augsburg.de}

\date{\today}

\begin{abstract}
We investigate transport through molecular wires whose energy levels
are affected by environmental fluctuations.  We assume that the
relevant fluctuations are so slow that they, within a tight-binding
description, can be described by disordered, Gaussian distributed
onsite energies.
For long wires, we find that the corresponding current distribution
can be rather broad even for a small energy variance.  Moreover, we
analyse with a Floquet master equation the interplay of laser
excitations and static disorder. Then the disorder leads to spatial
asymmetries such that the laser diving can induce a ratchet current.
\end{abstract}

\maketitle

\section{Introduction}

Chemical adsorption of sulfur atoms on gold surface allows a stable
bond between gold tips and thiol groups of molecules.  This has been
exploited for measuring the conductance and the current-voltage
characteristics of gold-molecule-gold junctions \cite{Datta1997a,
Reed1997a, Kergueris1999a, Cui2001a, Reichert2002a, Dadosh2005a}.  Repeated
measurements even with the same sample, however, reveal small but
noticeable differences which possibly stem from environmental
fluctuations that impact upon the effective molecule parameters.
Moreover, the particular form of the gold tip can have a significant
influence on the transport properties \cite{Yaliraki1998a}.

A present line of experimental research is the measurement of
molecular conductance when the electrons are excited by
electromagnetic waves.  There one expects various phenomena ranging
from photo-assisted transport \cite{Platero2004a, Kohler2005a} to
ratchet or non-adiabatic pump effects, i.e.\ the
induction of dc currents by ac fields even in the absence of any
voltage bias \cite{Lehmann2002b, Strass2005b}.  Moreover, it has been
predicted that properly taylored laser pulses can give rise to short
current pulses \cite{Franco2007a, Li2007a, Kohler2007a, Fainberg2007a}.
Since a dc current flows into one particular direction, a ratchet effect
can occur only in ``sufficiently asymmetric'' systems
\cite{Kohler2005a}.  In that respect, a static disorder is sufficient
to break the reflection symmetry of an individual realization and,
thus, may support a ratchet effect.

The quantitative prediction of the current through a molecule is still
a great challenge despite the significant progress achieved in
recent years \cite{DiVentra2000a, Damle2002a, Heurich2002a,
Evers2004a}.  For a more qualitative understanding of
the mechanisms involved in molecular transport, it is thus
advantageous to employ for the molecule a rather generic tight-binding
model \cite{Mujica1994a, Segal2000a, FoaTorres2001a, Cizek2004a,
Platero2004a, delValle2005a, Kohler2005a, delValle2007a}.
Then a flexible method for the computation of transport properties is
provided by master equations of the Bloch-Redfield type which allow
one to include electron-electron and electron-phonon interactions, as
well as time-dependent fields \cite{Kohler2005a, Kaiser2006b}.
Similar methods have also been used for describing incoherent
transport \cite{Petrov2001a, Lehmann2002a}.

Here, we explore the role of slow fluctuations or static disorder for
molecular conductance.  Thereby we will assume that the relevant
environmental fluctuations are so slow that they can be described as
static disorder which defines an ensemble of wire Hamiltonians.  Then
a natural quantity of interest is the corresponding distribution of
stationary currents.  A setup for which this current distribution is
also directly relevant is an array of molecular junctions that conduct
in parallel.
We employ a tight-binding model for the molecule and treat it with the
Floquet master equation formalism derived in Ref.~\cite{Kaiser2006b}
which we review briefly in Section~\ref{sec:model}.
In Section~\ref{sec:static}, we present results for
a static model with a large voltage bias, while in
Section~\ref{sec:driven}, we investigate pumping effects caused by an
interplay of ac driving fields and disorder. The analytical derivation
of the current distribution for a wire with two sites is deferred to
the Appendix.

\section{Wire-lead model and master equation}
\label{sec:model}

\begin{figure}[tb]
\centerline{\includegraphics{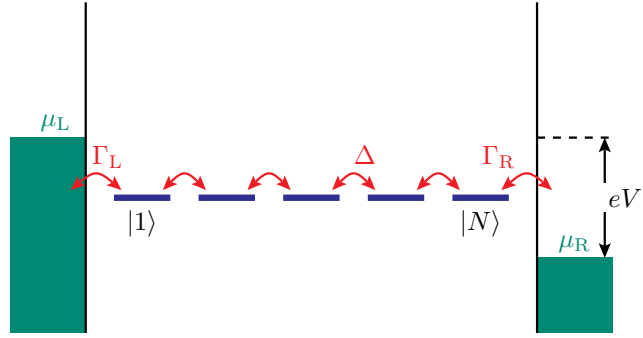}}
\caption{Bridged molecular wire model consisting of $N=5$ sites with internal
tunnelling matrix elements $\Delta$ and effective wire-lead coupling
strengths $\Gamma_\mathrm{L/R}$.}
\label{fig:wire_model}
\end{figure}
The system of the driven molecular wire, the leads, and the coupling
between the molecule and the leads, as sketched in
Fig.~\ref{fig:wire_model}, is described by the Hamiltonian
    \begin{equation}
    \label{eq:full_hamiltonian}
    \mathcal{H}(t) = \mathcal{H}_{\mathrm{wire}}(t) + \mathcal{H}_{\mathrm{leads}} +
    \mathcal{H}_{\mathrm{wire-leads}}.
    \end{equation}
The wire is modelled by $N$ tight-binding orbitals $\ket{n}$,
$n=1,\ldots,N$, such that
\begin{equation}
  \label{eq:wire_hamiltonian}
  \mathcal{H}_{\mathrm{wire}} =
      \sum_{n} (E_{n}(t)+\xi_n) c_n^\dagger c_{n}
    - \Delta \sum_{n=1}^{N-1} (c_{n+1}^\dagger c_{n} +
      c_{n}^\dagger c_{n+1})
    + \frac{U}{2} \mathcal{N}(\mathcal{N}-1) ,
\end{equation}
with the tunnel matrix element $\Delta$ and the capacitive energy $U$.
Each onsite energy $E_n(t)+\xi_n$ contains a random contribution $\xi_n$
that subsumes the influence of environmental fluctuations.  We assume
that these fluctuations are Gaussian distributed and so slow that we
can treat them as static disorder.  Thus, the probability that the
onsite energy of orbital $n$ lies in an interval of size
$\mathrm{d}\xi$ around $E_n(t)+\xi_n$ reads
\begin{equation}
\label{xi}
w(\xi_n) =
\frac{1}{\sqrt{2\pi\sigma^2}}\exp\Big(-\frac{\xi_n^2}{2\sigma^2}\Big) ,
\end{equation}
where the variance $\sigma^2$ is assumed to be position-independent.  This
implies that the energy fluctuations are spatially uncorrelated, such
that $\langle\xi_n\xi_{n'}\rangle = \sigma^2 \delta_{nn'}$. The
onsite energies $E_n (t)= E_n + A x_n \cos(\Omega t)$ are modulated
by a harmonically time-dependent dipole force, where $A$ denotes
the electrical field amplitude multiplied by the electron charge and the
distance between neighbouring sites, with $x_n = \frac{1}{2}(N+1-2n)$
the scaled position of site $|n\rangle$.
Our goal will be to compute for many realizations of the wire
Hamiltonian the resulting dc current which provides the current
distribution $P(I)$.
The last term in Eq.~\eref{eq:wire_hamiltonian} captures the electron-electron
interaction within a capacitor model and the operator $\mathcal{N}
= \sum_{n} c_{n}^\dagger c_{n}$ describes the number of excess
electrons residing on the molecule.  Below we shall assume that $U$ is
so large that only states with zero or one excess electron play a
role.   
The first and the last site of the molecule, $\ket{1}$ and $\ket{N}$,
couple via the tunnelling Hamiltonian
    \begin{equation}
    \label{eq:tunnel_hamiltonian}
    \mathcal{H}_{\mathrm{wire-leads}} =  \sum_{q} (V_{\L q}\,
    c^{\dagger}_{\L q}
    c_{1}^{\vphantom{\dagger}} + V_{\R q}\, c^{\dagger}_{\R q}
    c_{N}^{\vphantom{\dagger}}) + \mathrm{H.c.}
    \end{equation}
to the respective lead.  The operator $c^{\dagger}_{\L q}$
($c^{\dagger}_{\R q}$) creates an electron in the left (right) lead
in state $|\L q\rangle$ which is orthogonal to all wire states.
The influence of the tunnelling Hamiltonian is fully characterised by the
spectral density
$  \label{spectral.density}
  \Gamma_\ell(\epsilon) =
         2\pi\sum_q |V_{\ell q}|^2 \delta(\epsilon-\epsilon_q)$.
If the lead states are dense and located at the centre of the
conduction band, the spectral densities can be replaced by a constant,
i.e.\ we assume $\Gamma_\ell(\epsilon) = \Gamma$ for both leads.

The leads are modelled as ideal Fermi gases
    \begin{equation}
    \label{eq:lead_hamiltonian}
    \mathcal{H}_{\mathrm{leads}} =  \sum_{q} \big(
    \epsilon^{\vphantom{\dagger}}_{q\L}
    \, c^{\dagger}_{\L q} c^{\vphantom{\dagger}}_{\L q}+
    \epsilon^{\vphantom{\dagger}}_{q\R}
    \, c^{\dagger}_{\R q} c^{\vphantom{\dagger}}_{\R q}
    \big) ,
    \end{equation}
which are initially at thermal equilibrium with the chemical potential
$\mu_{\L/\R}$ and, thus, are described by the density operator
$    \rho_{\mathrm{leads,eq}} \propto
    \exp{[-(\mathcal{H}_{\mathrm{leads}}-\mu_\L \mathcal{N}_\L-\mu_\R
    \mathcal{N}_\R)/k_\mathrm{B}T]},$
where $\mathcal{N}_\ell = \sum_q c_{q\ell}^\dagger c_{q\ell}$ is the
electron number operator for lead $\ell = \L,\R$.  
Since a typical metal screens all electric fields with a frequency below the
plasma frequency, we assume that the bulk properties of the leads are
not affected by the laser irradiation.

\subsection{Perturbation theory}

The derivation of a master
equation starts from the Liouville-von Neumann equation $\i\hbar \dot
\rho(t)=[\mathcal{H}(t),\rho(t)]$ for the total density operator $\rho(t)$ for
which one obtains by standard techniques the approximate equation of
motion \cite{Bruder1994a, Gurvitz1996a, Lehmann2004a, Kohler2005a,
Brandes2005a}
\begin{equation}
\fl
\label{mastereq-gen}
  \dot\rho(t)
  =  -\frac{\i}{\hbar}[H_\mathrm{wire}(t)+H_\mathrm{leads},\rho(t)] 
     -\frac{1}{\hbar^2}\int_0^\infty \d\tau
      [H_\mathrm{wire-leads},[\widetilde
      H_\mathrm{wire-leads}(t-\tau,t),\rho(t)]] . \nonumber
\end{equation}
Here the first term corresponds to the coherent dynamics of both the wire
electrons and the lead electrons, while the second term describes
resonant electron tunnelling between the leads and the wire.
The tilde denotes operators in the interaction picture with respect to
the molecule and the lead Hamiltonian without the molecule-lead
coupling, $\widetilde X(t,t')=U_0^\dagger(t,t')\,X\,U_0(t,t')$, where $U_0$
is the propagator without the coupling.  The net (incoming minus
outgoing) electrical current through the left contact is given by
minus the time-derivative of the electron number in the left lead
multiplied by the electron charge $-e$.  From Eq.~\eref{mastereq-gen}
follows for the current in the wide-band limit the expression
\begin{eqnarray}
\fl
  I_\L(t)
  =  e \mathop{\mathrm{tr}}[\dot \rho(t) \mathcal{N}_\L] \nonumber
  =
   & -e\frac{\Gamma_\L}{\pi\hbar}\mathop{\mathrm{Re}}\int_0^\infty
   \d\tau \int\d\epsilon\, \e^{\i \epsilon \tau/\hbar}
   \label{current-general}\nonumber \\ 
    & \times
    \big\{ \langle c^\dagger_1 \tilde{c}_1^{\vphantom{\dagger}}
    (t,t-\tau)\rangle 
    \bar{f}_\L (\epsilon) -
    \langle \tilde{c}_1^{\vphantom{\dagger}}(t,t-\tau)c_1^\dagger
    \rangle 
    f_\L(\epsilon)\big\} ,
\end{eqnarray}
where $f_\ell$ is the Fermi function of the respective lead and $\bar f_\ell =
1-f_\ell$. Furthermore, $\langle\cdots\rangle =
\mathop{\mathrm{tr}_\mathrm{wire}}\rho_\mathrm{wire}\cdots$ denotes the
expectation value with respect to the wire density operator.
We emphasise that the expectation values in
Eq.~\eref{current-general} depend directly on the dynamics of the
isolated wire and are thus influenced by the driving.

\subsection{Floquet theory}

An important feature of the current formula \eref{current-general} is its
dependence on solely the wire operators while all lead operators have
been eliminated.  Therefore it is sufficient to consider the reduced
density operator $\rho_\mathrm{wire} =
\mathop{\mathrm{tr}_\mathrm{leads}}\rho$ of the wire electrons.  
The effort necessary to calculate $\rho_\mathrm{wire}$ can be reduced
significantly by exploiting the fact that
the master equation \eref{mastereq-gen} inherited from the total
Hamiltonian $\mathcal{H}(t)$ a periodic time-dependence, which opens the way for
a Floquet treatment.

One possibility for such a treatment is to use the Floquet states of
the central system, i.e.\ the driven wire, as a basis.  Thereby we
also use the fact that in the wire
Hamiltonian~\eref{eq:wire_hamiltonian}, the single-particle
contribution commutes with the interaction term and, thus, these two
Hamiltonians possess a complete set of common eigenstates.  
In analogy to the quasimomenta in Bloch theory for spatially periodic
potentials, the quasienergies $\epsilon_{\alpha}$ come in classes
$    \epsilon_{\alpha,k}=\epsilon_\alpha + k\hbar \Omega,
    k \in  \mathbb{Z},$
of which all members represent the same physical solution of the Schr\"odinger
equation. Thus we can restrict ourselves to states within one Brillouin
zone like for example $0 \leq \epsilon_\alpha < \hbar \Omega$.

For the computation of the current it is convenient to have an
explicit expression for the interaction picture representation of the
wire operators.  It can be obtained from the (fermionic) Floquet creation
and annihilation operators defined via the
transformation
$    \label{c.alpha}
    c_{\alpha } (t) = \sum_n \bra{\varphi_{\alpha}(t)} n\rangle c_{n}, $
which reads in the interaction picture 
$  \tilde{c}_{\alpha}(t,t')
  = \exponent{- \i(\epsilon_{\alpha}+U \mathcal{N}_\mathrm{wire}) (t-t') / \hbar}
    c_{\alpha}(t') ,$
with the important feature that the time difference $t-t'$ enters
only via the exponential prefactor \cite{Kohler2005a}.

\subsection{Master equation and current formula}

In the following, we assume the interaction $U$ to be the dominant
energy scale in the system, therefore we can restrict the wire Hilbert
space to the $N+1$ dimensional subspace of states with zero and one
electron, such that a basis for the decomposition of the
reduced operator is $\{ |0\rangle, c_{\alpha }^\dagger(t)\,|0\rangle \}$,
where $|0\rangle$ denotes the wire state in the absence of excess electrons.
Moreover, it can be shown \cite{Kaiser2006b} that at large times, the
density operator of the wire becomes diagonal in the electron number
$\mathcal{N}$.  Therefore a proper ansatz reads
\begin{equation}
  \label{eq:decomposition}
  \rho_{\mathrm{wire}}(t)
  = |0\rangle \rho_{00}(t) \langle 0|
    + \sum_{\alpha,\beta} c^\dagger_{\alpha }
    \ket{0}\rho_{\alpha \beta }(t)\bra{0}c_{\beta } .
\end{equation}
Note that we keep terms with $\alpha\neq\beta$, which means that we
work beyond a rotating-wave approximation.  

Following our evaluation of the master equation \cite{Kaiser2006b}, we arrive at
a set of $N^2$ coupled equations of motion for
$\rho_{\alpha\beta}(t)$ which in Fourier representation read
\begin{eqnarray}
\label{eq:masterfourier}
\fl
    \i(  \epsilon_{\alpha}-\epsilon_{\beta}-k\hbar\Omega)
    \rho_{\alpha \beta,k} \nonumber 
   &=& \frac{\Gamma_\L }{2}\sum_{k',k''}\,
    \langle \varphi_{\alpha,k'+k''} | 1 \rangle 
    \langle 1 | \varphi_{\beta, k+k''} \rangle
    \rho_{00,k'}
    \big( f_\L(\epsilon_{\alpha,k'+k''})
    +f_\L(\epsilon_{\beta,k+k''}) \big)\nonumber
    \\
    &&-\frac{\Gamma_\L }{2}\sum_{\alpha',k',k''}
    \langle \varphi_{\alpha,k'+k''} | 1 \rangle
    \langle 1 | \varphi_{\alpha',k+k''} \rangle
    \rho_{\alpha'\beta, k'}\,
    \bar{f}_\L (\epsilon_{\alpha',k+k''})\nonumber
    \\
    &&-\frac{\Gamma_\L }{2}\sum_{\beta',k',k''}
    \langle \varphi_{\beta',k'+k''} | 1 \rangle
    \langle 1 | \varphi_{\beta,k+k''} \rangle
    \rho_{\alpha\beta', k'}\,
    \bar{f}_\L (\epsilon_{\beta',k'+k''})\nonumber
    \\
    &&+\mathrm{same\ terms\ with\ the\ replacement}\:
    1, \L \rightarrow N, \R.
\end{eqnarray}
In an analogous manner we obtain for the dc current the expression
\begin{eqnarray}
\label{eq:dc-current}
    I_{\L} &= &\frac{e \Gamma_\L}{\hbar} \Re \sum_{\alpha,k}\Big(
    \sum_{\beta,k'}
    \langle \varphi_{\beta,k'+k} | 1 \rangle
    \langle 1 | \varphi_{\alpha,k} \rangle
    \rho_{\alpha \beta, k'}
    \bar{f}_\L(\epsilon_{\alpha,k})\nonumber
    \\ &&- 
    \sum_{k'}
    \langle \varphi_{\alpha,k'+k} | 1 \rangle
    \langle 1 | \varphi_{\alpha,k} \rangle 
    \rho_{00,k'}
    f_\L(\epsilon_{\alpha,k})
    \Big).
\end{eqnarray}

The results of this section allow us to numerically compute the dc
current through a driven conductor as well as studying the undriven
limit.  The current distributions discussed below are obtained by
computing the dc current for typically $10^4$ realizations of the wire
Hamiltonian \eref{eq:wire_hamiltonian}.  Then these values are taken for
a histogram with 150 bins which finally will be scaled such
that we obtain a normalised probability density.

\section{Electron transport with slowly fluctuating energies}
\label{sec:static}

We first address an undriven wire in the configuration sketched in
Fig.~\ref{fig:wire_model} where the distribution of all wire
levels is centred at energy $E_n=0$.  The transport voltage is so
large that all eigenenergies lie well within the voltage window and,
consequently, the transport is unidirectional.
Then in the absence of onsite energy fluctuations ($\sigma=0$), the
current can be evaluated analytically within a rotating-wave
approximation and reads $I_\mathrm{max}=e\Gamma/\hbar (N+1)$, i.e.\
it decays with increasing wire length \cite{Kaiser2006a}.
The index ``max'' refers to the fact that any modification of the
onsite energies can only reduce the current---which is confirmed
by our simulations.  The physical reason for this is that for equal
onsite energies, solely the kinetic energy determines the eigenstates
which, consequently, are delocalised.  Different onsite energies, by
contrast, tend to ``localise'' the eigenstates.
Thus in the limit of small disorder, the current distribution $P(I)$
is expect to possess a clear peak at $I=I_\mathrm{max}$ and some minor
contribution for lower values of $I$.

\begin{figure}
 \centerline{\includegraphics{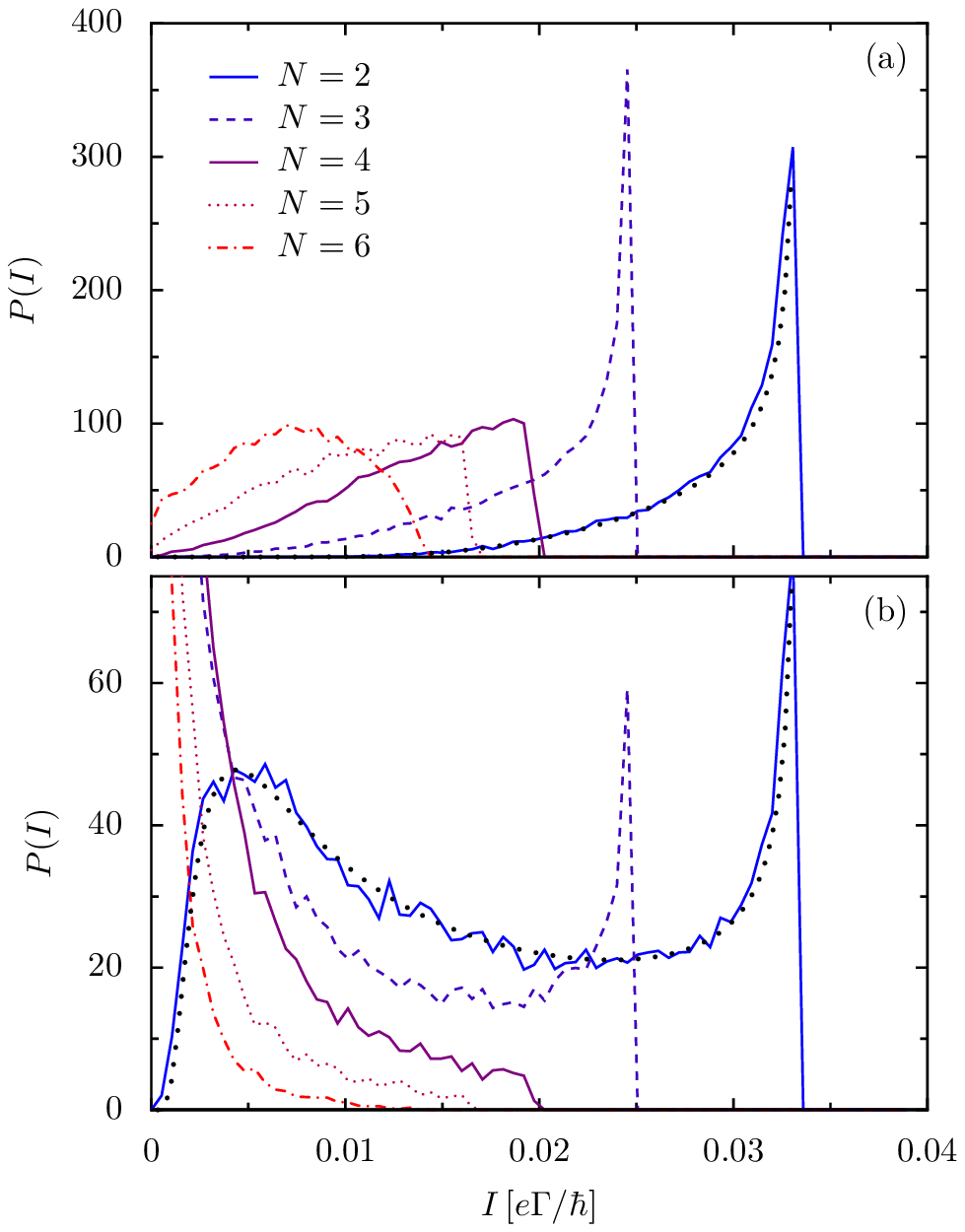}}
 \caption{Current distribution for a channel with $N$ sites in the
   limit of a large bias voltage.  The standard deviation of the onsite
   energies is $\sigma=0.5\Delta$ (a) and $\sigma=2\Delta$ (b),
   while the wire-lead coupling is $\Gamma=0.1\Delta$.  The distributions
   have been obtained by computing the current for $1.5\times 10^4$
   realizations of the wire Hamiltonian.  The black dotted lines mark
   the analytical results for $N=2$ sites.}
 \label{fig:static}
\end{figure}

Figure~\ref{fig:static} shows the simulated current distributions for
two different variances.  For a small variance (panel a), the
distributions for short wires show the expected behaviour.
With an increasing wire length, the peak at $I=I_\mathrm{max}$
disappears and is eventually replaced by an apparently parabolic distribution.
This length dependence can be understood in the following way: For a
short wire, the probability that a level is out of resonance is rather
low and, thus, most realizations of the wire Hamiltonian will allow
resonant inter-site tunnelling. With an increasing number of levels,
however, the probability for having at least one misaligned level
increases and a current significantly smaller than $I_\mathrm{max}$
becomes more likely. The precise values will depend on the details
and, consequently, we expect a broad distribution.  This means that
whenever a large number of levels plays a role, the transport through
a molecule is extremely sensitive to even small disorder induced
by environmental fluctuations.

With a larger variance, this scenario becomes even more pronounced as can be
seen in Fig.~\ref{fig:static}b: Then the peak at $I_\mathrm{max}$ is
rather small and noticeable only for 2 and 3 sites.  The most likely
realization is a completely disordered wire with an accordingly low
conductance.  For $N>3$, the distributions even possesse a significant
peak at $I=0$ which corresponds to isolating behaviour. A closer
inspection of the numerical data reveals that the crossover between
conducting and isolating behaviour occurs when the effective disorder
$\sqrt{N}\sigma$ exceeds the tunnel matrix element $\Delta$.

Interestingly enough, for $N=2,3$ the distribution turns out to be
even non-monotonic, which means that most one most likely finds
either a current close to the theoretical maximum or a significantly
smaller lower value.  The non-monotonic behaviour for $N=2$ can also be
found analytically. The derivation of the corresponding current
distribution~\eref{P.tls} can be found in the Appendix.
The excellent agreement of this analytical solution and the simulated
distributions emphasis that the simulation with approximately $10^4$
realisations ensures good convergence.

\section{AC-driven disordered junctions}
\label{sec:driven}

\begin{figure} 
 \begin{center}
    \includegraphics{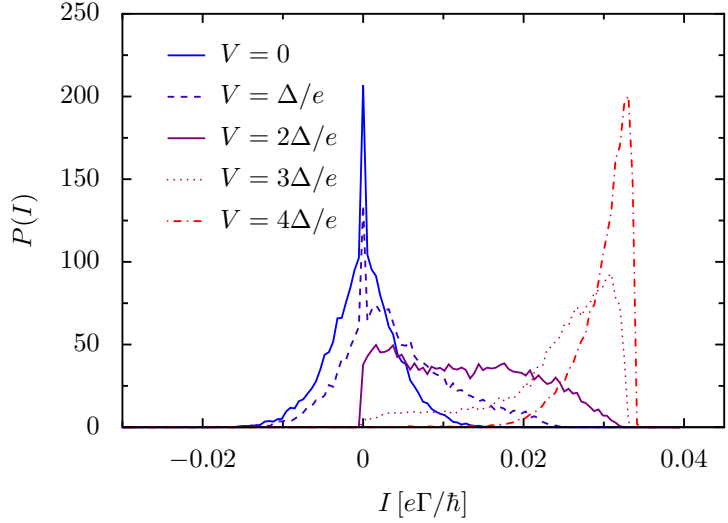} 
 \end{center} 
 \caption{Current distribution for an AC driven wire with $N=2$ sites
 for various bias voltages.  The fluctuations of the onsite
 energies are characterised by the standard deviation $\sigma=0.5\Delta$,
 the driving frequency and amplitude are $A=\Delta$ and
 $\Omega=2\Delta/\hbar$, respectively.
 All the other parameters are as in Fig.~\ref{fig:static}.}
 \label{fig:driven_system} 
\end{figure}

In order to investigate the influence of an AC driving, we employ the
same model as above, but now with an additional dipole driving
modelled by time-dependent onsite energies $E_n(t)=Ax_n\cos(\Omega t)$
as discussed in Sect.~\ref{sec:model}.  The driving frequency $\Omega
=2 \Delta/\hbar$ is chosen such that it matches the average splitting
of the wire energies, while the amplitude $A=\Delta$ corresponds to
intermediately strong driving.  The solid line in
Fig.~\ref{fig:driven_system} shows the current distribution in the
absence of a voltage bias, $V=0$.  The reflection symmetry of the ensemble
relates to the symmetric shape of the distribution, which implies
that the current vanishes in the ensemble average.  An individual
realization of the wire, however, generally does not possess reflection
symmetry because the random energy shifts are spatially uncorrelated.
This asymmetry in combination with the non-adiabatic driving
induces a coherent ratchet current, i.e.\ a dc current even in the
absence of any net voltage bias.  In the present case,
the ratchet current is of the order of 10--20 percent of the current observed
above in the large bias limit.  This order of magnitude is typical
when the driving frequency or a multiple of the driving frequency lies
close to an internal resonance, while the intensity is moderate
\cite{Kaiser2006b}.  In addition to the broad distribution of ratchet
currents, $P(I)$ exhibits a peak at $I=0$.  This stems from
realizations for which the driving is well out of resonance.

For a bias voltage $V>0$, the ensemble no longer possesses reflection
symmetry and the current distribution is shifted towards positive
values (see Fig.~\ref{fig:driven_system}).  For sufficiently small
voltages $V\lesssim \Delta/e$, non-adiabatic pumping against
the voltage bias is still possible.  Rather surprisingly, the peak at
zero current remains.  It now corresponds to realizations for which
on the one hand, the driving is off-resonant while on the other hand,
both levels lie outside the voltage window.  With an increasing bias
voltage, the second condition is less frequently fulfilled, and
eventually the distribution assumes a form similar to that found for a
large voltage in the absence of driving.  Already for $V\approx
4\Delta/e$, the distribution is hardly distinguishable from the one
shown in Fig.~\ref{fig:static}a for a wire with $N=2$ sites in the
absence of driving.

\section{Conclusions}

We have investigated the current through a molecular wire with
disordered onsite energies. Such a disorder can stem from the
interaction with slow fluctuations of background charges in the
substrate.  In particular, we computed the resulting current
distribution for two typical cases, namely an ``open transport
channel'' and a driven molecular wire for which random energy shifts
break reflection symmetry and, thus, the driving can induce a ratchet current.

The open transport channel is characterised by tight-binding levels
with equal onsite energies, such that any misalignment stems from the
disorder.  Its main consequence is that as soon as the standard
deviation of the onsite energies exceeds the tunnel matrix elements,
the current distribution no longer peaks only at a finite value, but
also at zero.  For longer wires, only the peak at zero current
remains.  This isolating behaviour resembles Anderson localisation which
is found for electrons in a one-dimensional disordered lattice
\cite{Anderson1958a}. Note however, that we here considered short wires
far from the scaling limit in which Anderson localisation is usually
studied.

Since the random energy shifts break reflection symmetry, driving the
molecular wire with a laser field induces ratchet currents for which we
found a relatively broad distribution.  If the driving frequency is
far from any molecular excitation energy, the ratchet current
will be rather small, and we indeed found that this is the case for
very many wire realizations. It has the consequence that the
corresponding distribution possesses a spike at zero current.  This
means that non-adiabatic pumping of electrons against a voltage bias
becomes generally impossible whenever the relevant wire energy levels
lie well within the voltage window.

In conclusion, our results reveal that slow fluctuations or a static
disorder can have a significant effect on molecular conduction.  In
various cases, the current distribution emerges to be rather flat,
which means that even the magnitude of the current depends sensitively
on environmental influences.

\section*{Acknowledgements}

Financial support of the German Excellence Initiative
via the ``Nanosystems Initiative Munich (NIM)'' and of the
Elite Network of Bavaria via the International
Doctorate Program ``NanoBioTechnology'' is gratefully acknowledged.
This work has been supported by Deutsche
Forschungsgemeinschaft through SFB 484 and SPP 1243.

\appendix
\section{Analytical solution for two levels}
\label{sec:tls}

A wire model with $N=2$ sites represents an analytically solvable case
for which one observes a non-monotonic current distribution and
which can serve as test case for numerical implementations.  Here we
consider a two-level system with on-site energies $\xi_{1,2}$, i.e.\
with a bias $2\eta=\xi_1-\xi_2$.  Since the random energy shifts $\xi_n$ are
Gaussian distributed with variance $\sigma^2$, the bias $2\eta$ is
also Gaussian distributed but with variance $2\sigma^2$, i.e.\ its
distribution function reads
$w(\eta)=\exp(-\eta^2/\sigma^2)/\sqrt{\pi\sigma^2}$.

For the computation of the current, we restrict ourselves to the limit
of a large transport voltage such that both eigenenergies of the
two-level system lie within the voltage window.  Then, the Fermi
functions of the left and the right lead effectively become
$f_\mathrm{L}=1$ and $f_\mathrm{R}=0$.  In this case, transport can be
described within rotating-wave approximation (RWA) which practically
means that the reduced density operator of the wire is diagonal in
energy representation \cite{Kaiser2006a}.
Within RWA thus follows from the master equation~\eref{eq:masterfourier}
the occupation probability $\rho_{\alpha\alpha} = w_\alpha^1
/w_\alpha^2$ and, thus, $\rho_{00}= 1- \sum_\alpha w_\alpha^1
/w_\alpha^2$. The coefficients $w_\alpha^n = |\langle\phi_\alpha
|n\rangle|^2$ denote the overlap between the eigenstate
$|\phi_\alpha\rangle$ and the localised state $|n\rangle$ (Note that
in the undriven case, all non-vanishing contributions have sideband
index $k=0$, such that here the sideband index $k$ can be omitted).
Inserting this solution into the current formula~\eref{eq:dc-current},
we obtain $I=e\Gamma/\hbar(1+\sum_\alpha w_\alpha^1/w_\alpha^2)$.

The remaining task is now to diagonalise the single-particle
Hamiltonian which provides the coefficients $w_\alpha^n$.  For bias
$2\eta$ and tunnelling matrix element $\Delta$, the Hamiltonian in
pseudo-spin notation reads $H= \eta\sigma_z+\Delta\sigma_x$ and
possesses the eigenenergies $\pm\delta = \pm (\eta^2+\Delta^2)^{1/2}$.
The corresponding eigenvectors $\phi_\alpha$ are proportional to
$(\delta+\eta,\Delta)$ and $(\delta-\eta,\Delta)$, respectively, such
that $w_\alpha^1/w_\alpha^2 = (\delta\pm\eta)^2/\Delta^2$.  Then we
obtain for the current the expression
\begin{equation}
I(\eta) = \frac{e\Gamma}{\hbar} \frac{1}{3+4\eta^2/\Delta^2}
= \frac{I_\mathrm{max}}{1+4\eta^2/3\Delta^2} ,
\end{equation}
which assumes its maximum $I_\mathrm{max}=e\Gamma/3\hbar$ in the
unbiased limit $\eta=0$.

The probability distribution for the current relates to $w(\eta)$ via
\begin{equation}
\label{P.transform}
P(I) = \sum_i w(\eta_i)\Big|\frac{\mathrm{d}\eta_i}{\mathrm{d}I}\Big| ,
\end{equation}
where the summation considers all values of $\eta$ that fulfil the
condition $I=I(\eta)$.
After some straightforward algebra, we obtain by evaluating
expression~\eref{P.transform} the current distribution
\begin{equation}
\label{P.tls}
P(I) = \sqrt{\frac{3\Delta^2}{4\pi\sigma^2}}
\frac{I_\mathrm{max}/I^2}{\sqrt{I_\mathrm{max}/I-1}}
\exp\Big(-\frac{3\Delta^2}{4\sigma^2}(I_\mathrm{max}/I-1)\Big) ,
\end{equation}
which is defined and normalised on the interval $[0,I_\mathrm{max}]$.

\section*{References}
\bibliographystyle{iopart}


\end{document}